\documentclass[10pt]{iopart}
\usepackage{graphicx}
\usepackage{subfigure}
\usepackage{setstack}
\usepackage{iopams}
\expandafter\let\csname equation*\endcsname\relax
\expandafter\let\csname endequation*\endcsname\relax
\usepackage{todonotes}
\usepackage{cite}
\usepackage{lipsum}
\usepackage{hyperref}
\usepackage{amsmath}

\begin{document}
\ioptwocol[{
\title[IncepFormerNet: A multi-scale multi-head attention network for SSVEP classification]{IncepFormerNet: A multi-scale multi-head attention network for SSVEP classification}
\author{Yan Huang$^1$,Yongru Chen$^1$,Lei Cao\textsuperscript{1}\textsuperscript{,}\textsuperscript{*},Yongnian Cao$^2$,Xuechun Yang$^2$,Yilin Dong$^1$M,Tianyu Liu$^1$}
\address{School of Information Engineering, Shanghai Maritime University, Shanghai, China}
\address{Tiktok Inc, San Jose, CA, United States}
\begin{abstract}
In recent years, deep learning (DL) models have shown outstanding performance in EEG classification tasks, particularly in Steady-State Visually Evoked
Potential(SSVEP)-based Brain-Computer-Interfaces(BCI)systems. DL methods have been successfully applied to SSVEP-BCI. This study proposes a new model called IncepFormerNet, which is a hybrid of the Inception and Transformer architectures. IncepFormerNet adeptly extracts multi-scale temporal information from time series data using parallel convolution kernels of varying sizes, accurately capturing the subtle variations and critical features within SSVEP signals.Furthermore, the model integrates the multi-head attention mechanism from the Transformer architecture, which not only provides insights into global dependencies but also significantly enhances the understanding and representation of complex patterns.Additionally, it takes advantage of filter bank techniques to extract features based on the spectral characteristics of SSVEP data. To validate the effectiveness of the proposed model, we conducted experiments on two public datasets, . The experimental results show that IncepFormerNet achieves an accuracy of 87.41\% on Dataset 1 and 71.97\% on Dataset 2 using a 1.0-second time window. To further verify the superiority of the proposed model, we compared it with other deep learning models, and the results indicate that our method achieves significantly higher accuracy than the others.The source codes in this work are available at:
https://github.com/CECNL/SSVEP-DAN.\\
\end{abstract}
}]
\section{Introduction}
Brain-Computer Interface (BCI) is a communication or control system that allows real-time interaction between the human brain and external devices\cite{mak2009clinical}. BCI systems work by measuring brain signals that carry the user's intent and convert them into corresponding control signals for devices, enabling computer-based communication or the direct control of external devices\cite{jiang2020temporal}\cite{xu2021review}\cite{pei2021tensor}. EEG-based BCIs reflect brain intent through brainwave signals, and because of their convenience, low cost\cite{abiri2019comprehensive}and non-invasive nature, they have garnered widespread attention. After nearly 50 years of research, BCI technology has successfully transitioned from scientific demonstrations to experimental applications. It has been used in various fields, including healthcare, entertainment, and the military. Among the various EEG paradigms, the high signal-to-noise ratio and low training time of steady-state visual evoked potentials (SSVEP)\cite{rakshit2020hybrid}\cite{kumar2019designing}\cite{mao2020improve}make it one of the most popular.

Steady-state visual evoked potentials (SSVEP) are periodic rhythms generated by the modulation of cortical activity in response to visual stimuli presented at fixed frequencies\cite{zhu2010survey}. SSVEP primarily occurs in the occipital region of the cerebral cortex. In the design of SSVEP-based brain-computer interface (BCI) systems, different targets flicker at distinct frequencies, thereby eliciting specific SSVEP signals for each target. By analyzing the SSVEP signals from the visual cortex, the target on which the subject is focusing can be identified, enabling the brain to control external devices through SSVEP BCI. Therefore, accurately decoding SSVEP signals is crucial, because it allows for the identification of the specific frequency corresponding to the user's focus. Traditional decoding methods, such as Canonical Correlation Analysis (CCA)\cite{lin2006frequency}, are among the mainstream approaches in this field. To reduce the misclassification rate of EEG signals, Zhang and colleagues made further optimizations by proposing Multi-way Canonical Correlation Analysis (MwayCCA)\cite{zhang2011multiway} and L1-regularized MwayCCA (L1-MCCA)\cite{zhang2013l1}. Recently, Zhou introduced Multi-set Canonical Correlation Analysis (MsetCCA)\cite{zhang2014frequency}, which optimizes reference signals from the common features of multiple calibration trials without any artificial signals (i.e., sine or cosine signals), demonstrating better performance than MwayCCA and L1-MCCA\cite{gao2021improvement}. Given the characteristics of SSVEP signals, which exhibit clear harmonic components, researchers have developed Filter Bank Canonical Correlation Analysis (FBCCA)\cite{chen2015filter}, which explicitly combines the fundamental frequencies and harmonics of the stimulus frequencies to enhance classification performance. Nowadays, filter bank techniques are widely used, based on these methods, Extended Canonical Correlation Analysis (ECCA) and Single-template-based Canonical Correlation Analysis (ITCCA)\cite{bin2011high} have also been developed for SSVEP classification. Wang and colleagues proposed a novel SSVEP detection method based on TRCA spatial filtering\cite{nakanishi2017enhancing}, that significantly outperformed extended CCA in terms of classification accuracy and information transfer rate (ITR). Later, Liu and others introduced Task Discriminative Component Analysis (TDCA)\cite{liu2021improving}, which does not require ensemble techniques, to further improve the performance of SSVEP-BCI.

In recent years, deep learning (DL) has gained increasing attention\cite{krizhevsky2017imagenet}, and researchers have extensively developed and studyed the application of deep learning algorithms for BCI decoding\cite{zhou2018epileptic}. Deep learning models can be directly applied to raw data\cite{craik2019deep}, automatically extracting features from raw EEG signals for rapid decoding. Nowadays, deep learning models are widely used for EEG decoding and classification across various tasks. Researchers in SSVEP-BCI systems have begun exploring deep learning techniques to develop algorithms for recognizing SSVEP frequencies. Based on SSVEP data, models can generally be categorized into two types: those that use time-domain data as input and those that use frequency-domain data. Literature has introduced the EEGNet model\cite{lawhern2018eegnet}, a convolutional neural network specifically designed for processing EEG data, which accepts time-domain data as input to achieve rapid decoding of SSVEP signals. Similarly, a time-domain CNN model\cite{ding2021filter},, tCNN has been proposed, which employs filter bank techniques to enhance performance in short time windows (FB-tCNN), further improving decoding accuracy. Additionally, a hybrid network model has been proposed that combines convolutional neural networks (CNN) with long short-term memory (LSTM)\cite{pan2022efficient} networks was developed to enhance the decoding capability and generalization ability. Moreover, CCNN\cite{waytowich2018compact} utilizes frequency-domain data from SSVEP signals as input, as frequency-domain data is rich in amplitude and phase information. The introduction of CCNN indicates that spectral data is also beneficial for SSVEP classification. Chen and colleagues proposed an SSVEP decoding method based on an attention mechanism, the SSVEPformer model\cite{chen2023transformer}, which uses frequency-domain signals as input and incorporates filter bank techniques. An improved version, FB-SSVEPformer, further enhances network performance. Considering the topological relationships among EEG channels, a dynamic graph convolutional neural network (DDGCNN)\cite{zhang2024dynamic} has been proposed, innovatively utilizing graph models for SSVEP signal recognition.

Owing to the limitation imposed by the limited quantity of EEG data on the classification performance of many deep learning-based methods, improving classification performance within the available data poses a significant challenge. Therefore, this paper proposes a hybrid model based on Inception and Transformer\cite{luo2022data}\cite{aznan2019simulating}\cite{bassi2021transfer}. Inception is a deep convolutional neural network, a network structure based on multi-scale convolutions\cite{Szegedy_Ioffe_Vanhoucke_Alemi_2017}, aimed at addressing the issue of traditional CNNs when handling inputs of different sizes. Using convolution kernels of various sizes, features are captured at different scales. The traditional Inception structure\cite{szegedy2015going}, stacks multiple convolution kernels of different sizes and pooling layers in parallel, allowing the network to extract features at multiple scales. To avoid computational explosion, the Inception module typically applies a 1×1 convolution kernel to reduce the number of channels before using larger convolution kernels, thereby reducing the computational load. SSVEP signals exhibit periodic fluctuations at different frequencies and manifest as a mixture of multi-frequency oscillations in the time domain. Therefore, capturing key and significant features for classification tasks is particularly important. Inspired by the Inception structure, we use convolution kernels of different sizes to extract multi-scale features from the time-domain data, allowing the model to capture critical features of SSVEP signals at various scales, thereby improving classification accuracy. Transformer is one of the most promising model architectures. It was first used in machine translation and quickly became dominant in natural language processing owing to its outstanding performance\cite{devlin2018bert}. Subsequently, it was applied in the field of computer vision, achieving remarkable results\cite{dosovitskiy2020image}(Dosovitskiy et al., 2020). Transformer-based models are highly versatile. Considering the unique nature of SSVEP data, we utilize the Inception concept in combination with Transformer characteristics to decode SSVEP signals.

The deep neural network model proposed in this study, IncepFormerNet, is designed based on the characteristics of SSVEP signals, incorporating both temporal and spatial features of the data. In addition, it utilizes a filter bank technique to fully exploit the harmonic information in the signals, further enhancing classification performance. To validate the performance of the model, we used publicly available SSVEP datasets from Tsinghua University, specifically the Benchmark and BETA datasets, both of which contain 40 stimulation frequencies. In these datasets, we randomly selected 1-second segments of data. The proposed model achieved an average accuracy of 87.41\% on Dataset 1 and 67.73\% on Dataset 2.
\section{Methods}
‌\subsection{Dataset description}
In this section, two widely used public datasets, Benchmark and BETA, are introduced in detail. The public datasets are used to evaluate the performance of the proposed model.
By repeating this process,the local minima of the loss function can be determined thereby optimising the neural network. 
‌\subsubsection{Dataset 1\\}
Benchmark: This dataset was from a 40-target SSVEP-BCI speller, where all stimulation frequencies are encoded using the Joint Frequency and Phase Modulation (JFPM) method. The range of stimulation frequencies was 8 to 15.8 Hz, with an interval of 0.2 Hz. The phase range was 0 to 1.5$\pi$, with a step size of 0.5$\pi$. Data was collected using a 64-channel EEG recorder, and to reduce storage and computation, the data was downsampled to 250 Hz. The experiment involved 35 subjects, each performing 6 blocks, with each block containing 40 trials. Each trial lasted 6 seconds, including 0.5 seconds of visual cue, 5 seconds of stimulus flickering, and 0.5 seconds of screen blank time. The average visual latency across all subjects was 0.14 seconds. For more information, please refer to the Benchmark dataset literature\cite{wang2016benchmark}.
‌\subsubsection{Dataset 2\\}
BETA: This dataset shares the same frequencies and phases as the Benchmark dataset but differs in the arrangement of stimuli. Thus, The BETA stimulation paradigm is more suitable for practical applications. The dataset included 70 subjects, with each subject performing 4 blocks, and each block consists of 40 trials. For the first 15 subjects, each trial included 0.5 seconds of visual cue, 2 seconds of stimulus flickering, and 0.5 seconds of screen blank time. For the remaining 55 subjects, each trial consisted of 0.5 seconds of visual cue, 3 seconds of stimulus flickering, and 0.5 seconds of screen blank time. The average visual latency across all subjects was 0.13 seconds. For more information, please refer to the BETA dataset literature\cite{liu2020beta}.
‌\subsubsection{Data preprocessing\\}
This study focuses on analyzing steady-state visual evoked potential (SSVEP) data from nine channels in the occipital region selected from a total of 64 channels, including Pz, PO5, PO3, POz, PO4, PO6, O1, Oz, and O2 \cite{wang2016benchmark}\cite{liu2020beta}. To further select effective data segments, considering the visual delay of SSVEP, data segments from the stimulus onset, denoted as [Td, Td + Tw], were used. Here, Td represents the visual delay, which is 0.14 seconds for Dataset 1 and 0.13 seconds for Dataset 2, and Tw is the time window (Tw $\in$ {0.4 s, 0.5 s, …, 1.2 s}). Based on the characteristics of the amplitude information in SSVEP data, this paper employs three different ranges of filters to effectively extract the harmonic information from the data. Following the M3 method from the FBCCAchen2015filter, three sub-band filters were designed with frequency ranges of 6-50 Hz, 14-50 Hz, and 22-50 Hz to form a filter bank. Additionally, to mitigate overfitting during model training, 50 randomly selected continuous data points from the effective data were set to zero. For specific details, please refer to the cited literature\cite{wolpaw2002brain}.
‌\subsection{IncepFormerNet}
For the IncepFormerNet model proposed in this study, as shown in Figure.1, it illustrates the architecture, which is mainly divided into four modules: the channel fusion module, the temporal feature extraction module, the former module, and the classification module. The structure of the temporal feature extraction module is based on Inception model. It was modified according to the unique characteristics of SSVEP data to effectively extract features. The Transformer module is then employed to learn the temporal dependencies of the SSVEP data from a global perspective. Finally, a softmax layer is used for classification.
\begin{figure*}[ht]
    \centering
    \includegraphics[width=0.7\textwidth, trim=1cm 22cm 1cm 3cm, clip]{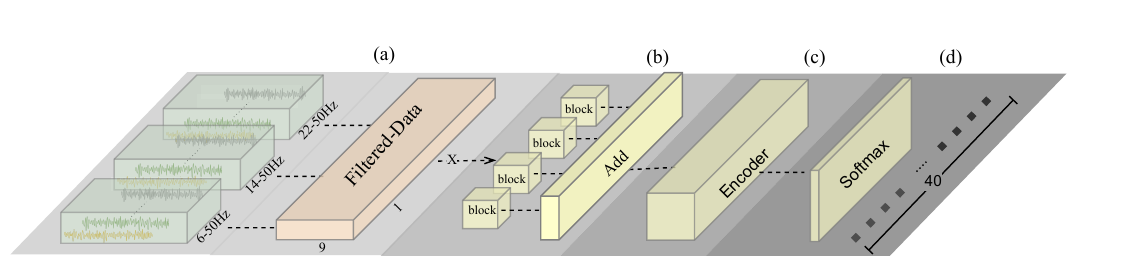}
    \caption{The diagram of IncepFormerNet model.(a) Channel Fusion Module.(b)Time Feature Extraction Module.(c)Former Module.(d)Classifier Module.}
\end{figure*}
‌\subsubsection{Channel Fusion Module\\}
The SSVEP data are derived from nine different channels, each containing EEG signals from different brain regions. Spatial
correlations may also exist between these channels. To fully utilize this information, convolution operations are used to apply weighted combinations to multiple channels\cite{lin2006frequency}. As shown in Figure.2, the convolutional layer uses a convolution kernel with a length of C×1 to assign different weights to each channel, generating a fused spatial feature. The formula for the channel fusion module is as follows, where information from different channels is fused while keeping the temporal dimension unchanged.
\begin{figure}[ht]
    \centering
    \includegraphics[width=0.45\textwidth,trim=1cm 2cm 1cm 2cm, clip]{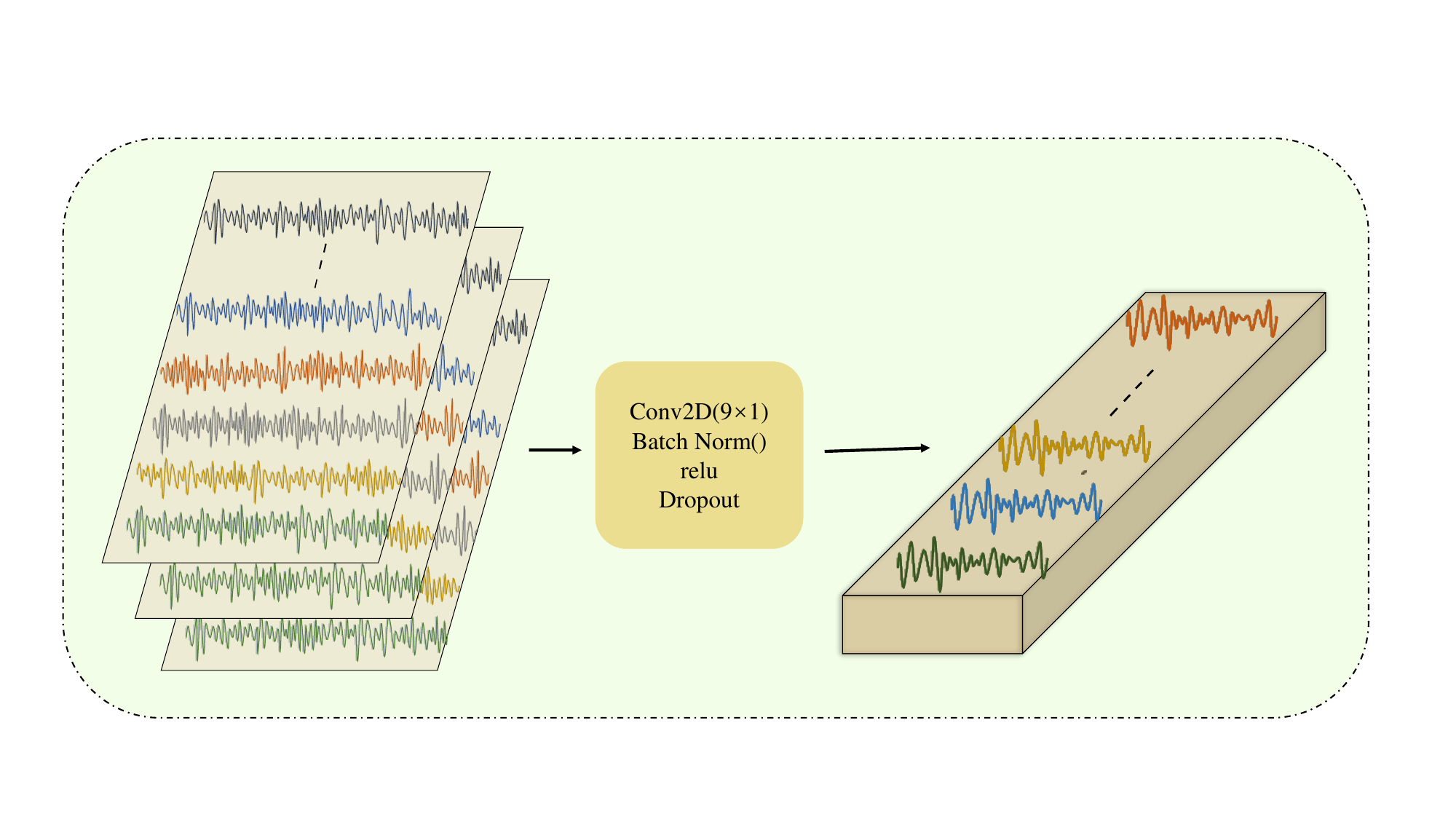}
    \caption{Apply convolution operations to the filtered data from three sub-bands to achieve a weighted combination across multiple channels, thereby generating a fused spatial feature.}
\end{figure}
\begin{equation}
Y_k (t) = \sum\nolimits_{i=1}^{G} W_k(i) \cdot X(i,t,b) \quad k=1,2,\ldots,9
\end{equation}
Where $W_k(i) $ represents the weight of the convolution kernel $W_k $ on channel i, $X $ represents the input data, and $Y_k (t) $ represents the value of the output feature at time point t and output channel k.
‌\subsubsection{Time Feature Extraction Module\\}
The temporal feature extraction module receives the data features after channel fusion. We apply four 1D convolutions and two pooling layers to the fused data for parallel computation. A larger convolution kernel is then used to further extract temporal information. Specifically,multi-scale convolutions are employed, combined with the concepts of pooling and residual connections, to extract signal features at different scales and layers,thereby enhancing prominent features. In this architecture, small convolution kernels focus on capturing local and prominent features of the SSVEP signals, while large convolution kernels capture longer-range signal features, handling features with temporal dependencies.After each convolution operation, Dropout and L2 regularization are applied to prevent overfitting and improve the model's generalization ability.Pooling layers are used to reduce the dimensionality of the feature maps, suppress noise interference,retain key prominent features, and enhance the model's robustness to interference.

The specific operations following the channel fusion of the data are shown in Figure.3. Inspired by\cite{szegedy2015going} the data are input into CNN blocks 1, 2, 3, and 4, resulting in the parallel convolution outputs $x_1,x_2,x_3$ and $x_4$. The output $x_2$ is then passed through CNN block 5 to obtain the convolution result $x_{\text{2\_2}}$ Subsequently, $x_{\text{2\_2}}$ and $x_3$ are concatenated and passed through CNN block 6 to obtain the result $x_{\text{3\_3}}$. Similarly, the result is concatenated with $x_4$ and passed through CNN block 7 to produce $x_{\text{4\_4}}$. Two pooling layers are concatenated to form $x_{\text{5\_5}}$. Finally, $x_2$, $x_{\text{2\_2}}$, $x_{\text{3\_3}}$, $x_{\text{4\_4}}$, and $x_{\text{5\_5}}$ are concatenated to obtain the temporal features of the signal.
\begin{figure*}[ht]
    \centering
    \includegraphics[width=0.8\textwidth,trim=1cm 17cm 5cm 2cm, clip]{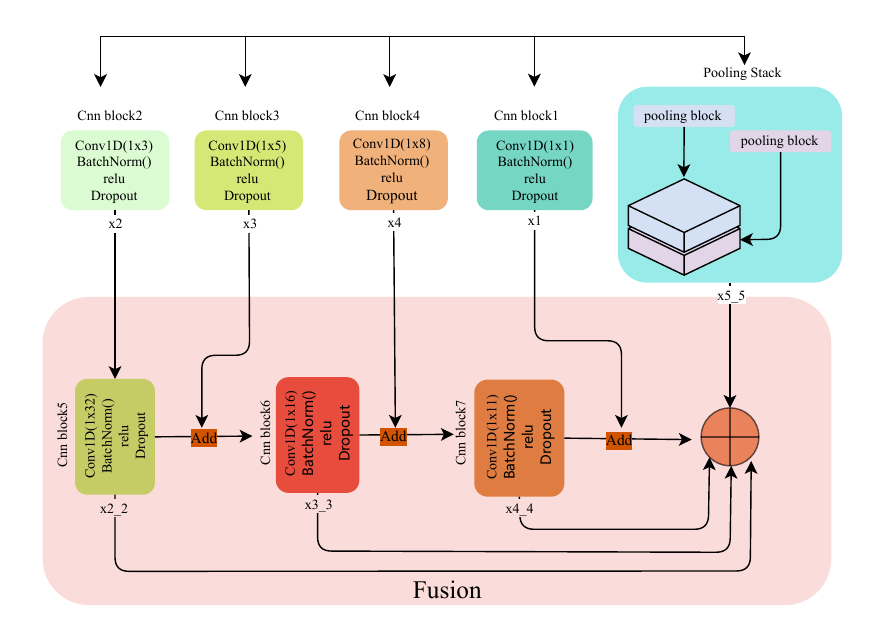}
    \caption{A detailed diagram of the temporal feature extraction module, which includes four convolutional kernels of different scales, with arrows indicating the respective fusion pathways.}
\end{figure*}
‌\subsubsection{Former Module\\}
The Transformer architecture consists of an encoder and a decoder, where the decoder is made up of multiple transformers with the same structure as described in "A Survey on Vision Transformer"\cite{han2022survey}. As shown in  Figure.4, the proposed  structure in this study only uses the encoder, with its primary task being feature extraction, ultimately leading to classification. The encoder layer receives the input data and processes the input sequence through multiple parallel attention heads, capturing different relationships within the sequence. The output of the multi-head attention is then connected to the input through a residual connection, followed by layer normalization. The main task of the feedforward neural network is to process the normalized data through two fully connected layers to extract features, followed by another residual connection and normalization. The final output is the features processed by the encoder. In this study, multiple encoder layers are stacked, with the output of each layer serving as the input to the next layer, allowing for the capture of complex temporal and global information by stacking multiple encoder layers.
In the attention mechanism, the core idea is to weight the relevance of the input data, allowing the model to pay attention to other data while processing a specific piece of information. The workflow is illustrated in the figure, where the input vector is first transformed into three different vectors: the query vector q, the key vector k, and the value vector v. These vectors from different inputs are then packaged into three distinct matrices, namely Q, K, and V\cite{han2022survey} which are used to calculate the attention scores. The calculation formula is as follows.
\begin{equation}
\text{Attention}(Q,K,V) = \text{softmax}\left(\frac{Q \cdot K^T}{\sqrt{d_k}}\right) \cdot V
\end{equation}
\begin{figure*}[ht]
    \centering
    \includegraphics[width=0.8\textwidth,trim=0.5cm 6cm 0.5cm 11.5cm, clip]{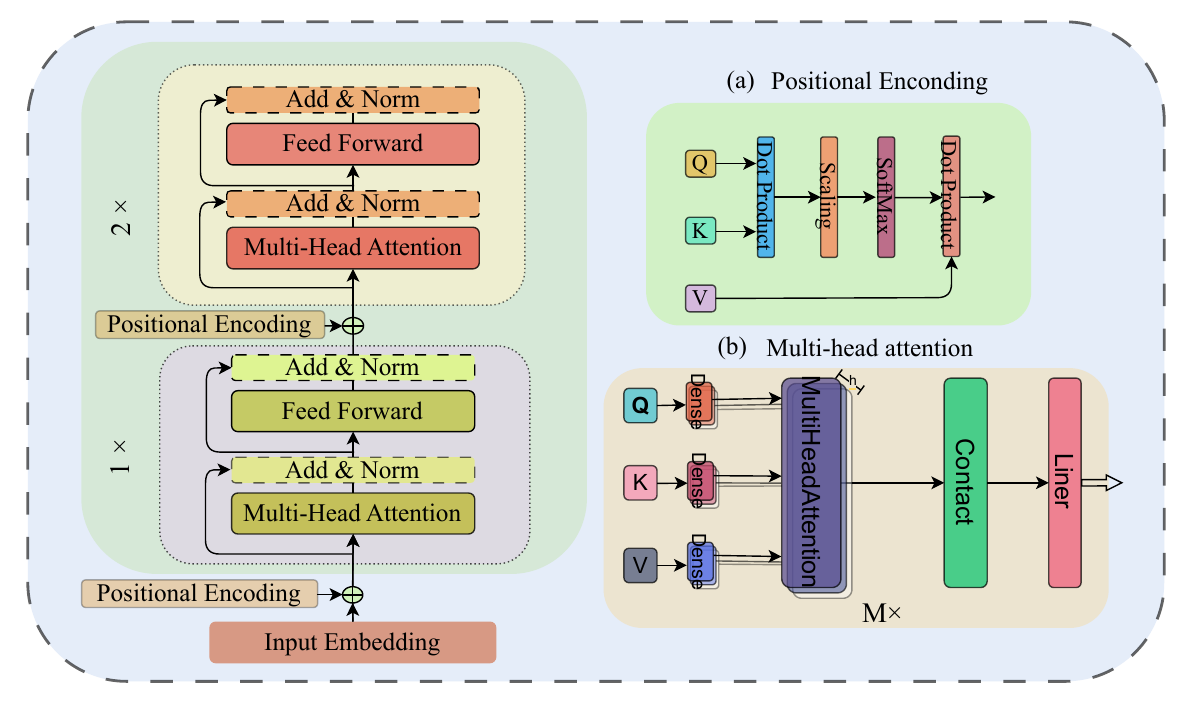}
    \caption{A detailed diagram of the Former Module, which employs a two-layer encoder. (a) Prefix encoding (b) Multi-head attention mechanism.}
\end{figure*}
The structure of multi-head attention mechanism is illustrated in Figure.4. It employs a set of linear transformation layers, in which three transformation tensors apply linear transformations to Q, K, and V. This design allows each attention mechanism to optimize different features of the data, enabling the model to learn multiple relationships of the data across different dimensions. In this paper, two encoder layers and four multi-head attention mechanisms are used.
‌\subsubsection{Classifier Module\\}
The classification module first uses a flattened layer to transform the multi-dimensional features into a one-dimensional vector, which is then inputted into a fully connected layer. This fully connected layer maps the features to the corresponding 40 output stimulus frequencies. Finally, a softmax activation function is applied to convert the output into a probability distribution of the classes, completing the classification of the SSVEP signals.
‌\subsection{Performance Evaluation}
Classification accuracy and information transfer rate are two important metrics for evaluating the performance of SSVEP classification methods and comprehensively assessing the system's performance from the perspectives of accuracy and time efficiency. These metrics play a crucial role in SSVEP research.
Classification accuracy is defined as the precision of detecting and classifying different stimulus frequencies, represented by the ratio of true positive samples to the total number of samples. A high classification accuracy indicates that the model can effectively capture and extract features from EEG signals, resulting in strong recognition capabilities and ensuring the robustness and usability.
Information transfer rate (ITR)\cite{wolpaw2002brain} is also crucial ("Improving the performance of individually calibrated SSVEP-BCI by Task-discriminant component analysis"). ITR combines the classification accuracy with the corresponding time required by the model. It is defined as:
\begin{equation}
    \text{ITR} = \frac{60}{T} \left[ \log_2 N + P \log_2 P + (1 - P) \log_2 \frac{1 - P}{N - 1} \right]
\end{equation}
Where N represents the number of target classes, P is the classification accuracy, and T is the selected sample data. By optimizing ITR, the accuracy of the system can be enhanced, leading to improved information transfer efficiency, thereby enhancing the practicality of SSVEP-BCI.
The the Benchmark dataset contained 35 subjects, each containing six blocks. A leave-one-out cross-validation method was used for the experiments. First, one block was randomly excluded to serve as the test set, whereas the remaining five blocks were used as the training set. This process was repeated six times to ensure that each block served as the test data with a unique training model. The model was trained using 5000 mini-batches for iterations. The test results for the six blocks were averaged to obtain the mean accuracy for each subject. Finally, the same procedure was applied to all 35 subjects to ensuring that each subject had a unique model, leading to the final test results for everyone. Similarly, in the BETA dataset, there were a total of 70 subjects, and the leave-one-out cross-validation method was employed to train models for each subject, ultimately yielding the average accuracy for each individual. In the BETA dataset, the number of mini-batches was set to 3000.

During the training process, dynamic learning rate adjustment was implemented by defining a LearningRateScheduler to prevent the learning rate from being too high and causing the model to fail to converge. An early stopping strategy was also incorporated to prevent overfitting during the training.
\subsection{The baseline methods}
‌\subsubsection{FBCCA\\}
FBCCA is based on the filter bank extension of CCA (Chen, Wang, Gao, et al., 2015). It decomposes EEG signals into sub-bands through a filter bank and performs CCA analysis. The final result is obtained by weighting the correlation coefficients from all the sub-bands.9
TRCA: TRCA is a spatial filtering method that extracts task-related components by maximizing the reproducibility of data in each task (Nakanishi et al., 2017). The average of existing data for the same task is then used as the reference signal for that task. Using the spatial filters obtained from TRCA based on calibration data, the correlation coefficients between the projected features of the test samples and various reference signals can be calculated to obtain the classification result. (In this study, for fairness, the calibration data selected is consistent with the test samples and the method presented in this paper)\cite{chen2015filter}.
‌\subsubsection{tCNN/FB-tCNN\\}
This is a time-domain-based CNN method. To avoid feature ambiguity in the frequency-domain paradigm within short time windows, the tCNN model was proposed. Considering that harmonic information embeds a large amount of effective information for frequency recognition, the tCNN was extended to FB-tCNN, which improves the network's classification performance within short time windows.\cite{ding2021filter}.
‌\subsubsection{EEGnet\\}
EEGNet is a convolutional neural network specifically designed for EEG signal data. Waytowich et al. applied EEGNet to SSVEP classification and achieved good results in cross-subject classification tasks (Waytowich et al., 2018)\cite{lawhern2018eegnet}.
‌\subsubsection{SSVEPformer\\}
Transformer-based models have been proven to be applicable to SSVEP-BCI. SSVEPformer uses frequency-domain signals as input and achieves impressive classification performance\cite{chen2023transformer}.
‌\subsubsection{CNNformer\\}
CNNformer combines CNN and Transformer. The CNN module captures temporal and spatial features, while the Transformer module learns global temporal dependencies, effectively improving the classification performance of SSVEP data\cite{ding2024novel}.
‌\subsubsection{DDGCNN\\}
DDGCNN is a dynamic graph convolutional neural network model that comprehensively considers the topological structure between channels. This model also addresses the over-smoothing problem in GCNs and is effective for learning and extracting EEG topological structure features\cite{zhang2024dynamic}.
\section{Result}
‌\subsection{Accuracy on two datasets}
Table 1 presents the classification accuracy of all subjects in the benchmark dataset under the proposed model. The time window was set to range from 0.6 seconds to 1.2 seconds, with an interval of 0.2 seconds, to gradually evaluate the model's performance across different time window lengths. Specifically, when the time window length is 0.6 seconds, the average classification accuracy for 35 subjects is 69.71\%. When the time window length is adjusted to 0.8 seconds, the average accuracy rises to 79.97\%. As the time window is further expanded, the accuracy continues to improve: with a window length of 1.0 seconds, the average accuracy reaches 87.41\%, and with a window length of 1.2 seconds, the average accuracy reaches 92.01\%. These results indicate that different time window settings significantly affect classification performance, with classification results varying according to the window length. Notably, in the 1.2-second time window, only one subject out of the 35 had an accuracy below 70\%, while the rest exceeded 70\%, achieving the benchmark for SSVEP-BCI, thereby validating the feasibility of the model. 
\begin{table*}[htbp]
\begin{center}
\caption{Accuracy of all subjects on the Dataset 1 with different data lengths.}
\begin{tabular}{c|cccc|c|cccc}
\hline
              & \multicolumn{4}{c|}{Data length(s)}   &                  & \multicolumn{4}{c}{Data length(s)}                                        \\ \cline{2-5} \cline{7-10} 
Subject(half) & 0.6     & 0.8     & 1.0     & 1.2     & Sbuject(half) & 0.6              & 0.8              & 1.0              & 1.2              \\ \hline
S1            & 76.80 & 87.86\ & 93.70 & 96.63 & S19              & 35.24          & 48.54          & 60.95          & 71.99          \\
S2            & 81.70 & 91.68\ & 96.99 & 98.75 & S20              & 69.75          & 83.72          & 92.54          & 96.28          \\
S3            & 85.60 & 93.43\ & 97.60 & 99.05 & S21              & 72.07          & 84.79          & 92.30          & 96.84          \\
S4            & 83.42 & 91.83\ & 95.20 & 97.74 & S22              & 85.57          & 93.58          & 97.15          & 98.86          \\
S5            & 81.88 & 91.17\ & 95.92 & 98.46 & S23              & 77.30        & 87.14          & 91.32        & 93.42          \\
S6            & 69.54 & 82.27\ & 91.92 & 96.50 & S24              & 73.58          & 84.81          & 92.43          & 95.45          \\
S7            & 64.01 & 78.74\ & 87.88 & 93.13 & S25              & 71.79          & 85.36          & 91.94          & 95.93          \\
S8            & 53.24 & 63.62\ & 74.83 & 84.77 & S26              & 77.18          & 87.28          & 93.77          & 96.53          \\
S9            & 57.83 & 67.27\ & 78.42 & 85.44 & S27              & 84.06          & 92.89          & 96.42          & 98.42          \\
S10           & 73.74 & 87.19\ & 94.46 & 97.58 & S28              & 73.91          & 85.58          & 92.79          & 95.90          \\
S11           & 38.03 & 50.02\ & 60.88 & 71.35 & S29              & 50.98          & 63.43          & 74.15          & 83.98          \\
S12           & 77.64 & 88.78\ & 94.76 & 97.58 & S30              & 70.10          & 80.02          & 88.56          & 93.24          \\
S13           & 73.54 & 85.86\ & 93.85 & 97.02 & S31              & 88.90          & 95.93          & 98.95          & 99.66          \\
S14           & 73.65 & 86.01\ & 93.45 & 96.60 & S32              & 85.47          & 94.58          & 98.31          & 99.28          \\
S15           & 51.84 & 67.22\ & 78.67 & 86.82 & S33              & 29.79          & 40.57          & 50.44          & 59.88          \\
S16           & 65.75 & 78.12\ & 88.42 & 93.94 & S34              & 65.76          & 78.51          & 87.05          & 91.91          \\
S17           & 69.93 & 83.21\ & 90.70 & 95.53 & S35              & 63.88          & 75.93          & 85.04          & 91.95          \\
S18           & 51.20 & 62.08\ & 67.71 & 74.10 & mean             & \textbf{68.71} & \textbf{79.97} & \textbf{87.41} & \textbf{92.01} \\ \hline
\end{tabular}
\end{center}
\end{table*}

Table 2 lists the accuracies of all subjects in the BETA dataset under the proposed model. Similarly, the time window was set to range from 0.6 seconds to 1.2 seconds, with an interval of 0.2 seconds. When the selected time window length is 0.6 seconds, the average accuracy for 70 subjects is 53.84\%. When the time window length is adjusted to 0.8 seconds, the average accuracy increases to 62.91\%. With a time window length of 1.0 seconds, the average accuracy reaches 67.73\%, and at a length of 1.2 seconds, the average accuracy is 71.97\%. Given that the BETA dataset has a shorter stimulation time and lower signal-to-noise ratio (SNR), achieving an average accuracy of 71.97\% is a relatively noteworthy result, which also validates the model's high classification performance. 
\begin{table*}[htbp]
\begin{center}
\caption{Accuracy of all subjects on Dataset 2 with different data lengths.}
\begin{tabular}{ccccc|ccccc}
\hline
\multicolumn{1}{c|}{}              & \multicolumn{4}{c|}{Data length(s)}                & \multicolumn{1}{c|}{}              & \multicolumn{4}{c}{Data length(s)}     \\ \cline{2-5} \cline{7-10}
\multicolumn{1}{c|}{Subject(half)} & 0.6                  & 0.8     & 1.0     & 1.2     & \multicolumn{1}{c|}{Subject(half)} & 0.6     & 0.8     & 1.0     & 1.2     \\ \hline
S1                                 & 66.40              & 78.52 & 82.99 & 89.23 & S36                                & 75.59 & 88.17 & 94.41 & 96.67 \\
S2                                 & 63.41              & 73.66 & 78.09 & 81.16 & S37                                & 81.69 & 92.94 & 95.75 & 97.75 \\
S3                                 & 58.82              & 70.02 & 76.11 & 81.95 & S38                                & 54.14 & 65.53 & 73.61 & 77.45 \\
S4                                 & 58.71              & 63.84 & 66.83 & 73.28 & S39                                & 40.04 & 48.28 & 57.74 & 64.04 \\
S5                                 & 62.01              & 72.25 & 73.24 & 82.26 & S40                                & 60.28 & 73.52 & 78.67 & 86.77 \\
S6                                 & 35.71              & 51.98 & 53.41 & 64.18 & S41                                & 26.74 & 35.86 & 34.68 & 36.36 \\
S7                                 & 24.37              & 30.35 & 34.47 & 29.43 & S42                                & 71.43 & 78.43 & 83.95 & 88.31 \\
S8                                 & 29.93              & 38.27 & 43.72 & 43.92 & S43                                & 60.83 & 68.45 & 74.41 & 79.92 \\
S9                                 & 65.59              & 75.97 & 81.11 & 86.03 & S44                                & 22.16 & 23.45 & 21.88 & 28.79 \\
S10                                & 44.31              & 44.85 & 51.38 & 56.08 & S45                                & 50.33 & 61.63 & 74.56 & 77.59 \\
S11                                & 12.81              & 13.76 & 14.94 & 15.15 & S46                                & 36.86 & 42.64 & 46.12 & 51.84 \\
S12                                & 78.08              & 87.18 & 90.33 & 95.23 & S47                                & 42.09 & 53.37 & 59.34 & 62.68 \\
S13                                & 76.64              & 85.01 & 90.51 & 93.66 & S48                                & 74.71 & 87.18 & 92.38 & 97.29 \\
S14                                & 43.59              & 53.25 & 58.19 & 49.19 & S49                                & 73.08 & 84.22 & 89.74 & 94.66 \\
S15                                & 55.73              & 63.04 & 69.42 & 73.04 & S50                                & 27.50 & 31.85 & 37.26 & 41.58 \\
S16                                & 46.03              & 54.67 & 64.57 & 69.67 & S51                                & 55.67 & 64.92 & 73.33 & 82.59 \\
S17                                & 9.75\              & 12.83 & 12.30 & 13.18 & S52                                & 67.04 & 77.03 & 85.53 & 90.71 \\
S18                                & 78.09              & 89.64 & 92.61 & 97.62 & S53                                & 52.64 & 60.83 & 70.77 & 71.14 \\
S19                                & 54.47              & 63.18 & 72.91 & 81.10 & S54                                & 35.60 & 44.15 & 46.83 & 55.76 \\
S20                                & 36.68              & 44.00 & 42.86 & 44.92 & S55                                & 17.16 & 20.36 & 22.71 & 25.48 \\
S21                                & 79.61              & 86.24 & 90.39 & 94.37 & S56                                & 70.54 & 82.88 & 90.50 & 93.96 \\
S22                                & 69.55              & 79.93 & 85.48 & 90.68 & S57                                & 76.91 & 87.13 & 91.68 & 95.55 \\
S23                                & 89.41              & 97.18 & 98.98 & 99.89 & S58                                & 73.91 & 85.19 & 90.79 & 92.69 \\
S24                                & 58.16              & 66.49 & 75.02 & 82.18 & S59                                & 18.63 & 25.67 & 27.18 & 31.49 \\
S25                                & 65.84              & 74.34 & 79.23 & 79.46 & S60                                & 50.01 & 63.16 & 70.38 & 77.41 \\
S26                                & 36.32              & 45.48 & 50.38 & 59.95 & S61                                & 5.83\ & 8.88\ & 9.90\ & 9.47\ \\
S27                                & 74.27              & 83.94 & 88.62 & 93.22 & S62                                & 55.55 & 63.74 & 70.01 & 80.38 \\
S28                                & 52.71              & 64.74 & 72.51 & 80.28 & S63                                & 78.56 & 89.15 & 93.92 & 97.88 \\
S29                                & 52.08              & 65.47 & 70.41 & 78.98 & S64                                & 66.52 & 79.43 & 87.44 & 91.74 \\
S30                                & 51.59              & 65.49 & 69.30 & 74.80 & S65                                & 18.76 & 25.17 & 21.83 & 26.50 \\
S31                                & 27.01              & 36.06 & 38.86 & 47.91 & S66                                & 78.18 & 86.01 & 89.26 & 90.06 \\
S32                                & 20.38              & 27.66 & 29.73 & 31.23 & S67                                & 87.53 & 94.26 & 97.12 & 98.68 \\
S33                                & 38.01              & 47.85 & 51.20 & 47.55 & S68                                & 77.67 & 90.01 & 94.31 & 97.40 \\
S34                                & 66.81              & 79.08 & 83.72 & 92.70 & S69                                & 65.67 & 77.26 & 86.95 & 91.74 \\
S35                                & 61.14              & 71.76 & 80.34 & 86.19 & S70                                & 75.00 & 85.11 & 92.31 & 95.89 \\
                               & \multicolumn{1}{l}{} &         &         &         &                   mean                         & \textbf{53.84} & \textbf{62.91} & \textbf{67.73} & \textbf{71.97} \\ \hline
\end{tabular}
\end{center}
\end{table*}

Table 3 presents the classification performance on the benchmark dataset for different models, comparing the average accuracies of FBCCA, TRCA, tCNN, SSVEPformer, CNNformer and DDGCNN across various time window lengths. It can be observed that the proposed model consistently outperforms all other models across all time windows, demonstrating its superiority in within-subject classification performance. This result validates the effectiveness of the proposed model, indicating its outstanding performance in classification tasks. This advantage is reflected not only in the improved accuracy but also in the adaptability of the model to different stimulation frequencies, demonstrating its stability and reliability when processing complex EEG signals.

Similarly, on the BETA dataset in Table 4, the proposed model exhibits strong advantages in comparison with CCA, FB-tCNN, EEGnet, and CNNformer. The table displays the average accuracies for different time window lengths. Across all time windows, the proposed model achieves the highest accuracy on the BETA dataset, further confirming the effectiveness of the model in diverse data environments. Through comparative analysis, it is evident that the proposed model maintains high classification performance across different time periods, reinforcing its potential and value in practical applications. This exceptional performance not only offers new insights for EEG signal classification but also lays a solid foundation for future related research.

\begin{table*}[htbp]
\begin{center}
\caption{Comparison of average accuracy across different methods and time windows on Dataset 1.}
\begin{tabular}{c|c|cccc}
\hline
                      &                            & \multicolumn{4}{c}{Data length(s)}                                \\ \cline{3-6} 
                      & methods                    & 0.6            & 0.8            & 1.0            & 1.2            \\ \hline
                      & FBCCA                      & 33.05          & 52.69          & 66.52          & 75.74          \\
                      & TRCA                       & 35.66          & 47.17          & 53.51          & 57.26          \\
                      & tCNN                       & 52.52          & 73.82          & 82.89          & 88.57          \\
Benchmark             & SSVEPformer                & 58.71          & 73.20          & 80.40          & 83.98          \\
\textbf{}             & CNNformer                  & 57.50          & 77.61          & 85.62          & 90.66          \\
\multicolumn{1}{l|}{} & DDGCNN                     & 64.80          & 74.70          & 83.20          & N              \\
\multicolumn{1}{l|}{} & \textbf{IncepFormer-SSVEP} & \textbf{68.71} & \textbf{79.97} & \textbf{87.41} & \textbf{92.01} \\ \hline
\end{tabular}
\end{center}
\end{table*}
\begin{table*}[htbp]
\begin{center}
\caption{Comparison of average accuracy across different methods and time windows on Dataset 2.}
\begin{tabular}{c|c|cccc}
\hline
          &                            & \multicolumn{4}{c}{Data length(s)}                                \\ \cline{3-6} 
          & methods                    & 0.6            & 0.8            & 1.0            & 1.2            \\ \hline
          & CCA                        & N              & 17.14          & 24.44          & 30.89          \\
Beta      & FB-tCNN                    & 42.59          & 49.20          & 54.07          & 57.70          \\
          & EEGnet                     & N              & 52.92          & 59.16          & 62.65          \\
          & CNNformer                  & 50.65          & 58.56          & 64.90          & 68.77          \\
\textbf{} & \textbf{IncepFormer-SSVEP} & \textbf{53.84} & \textbf{62.91} & \textbf{67.73} & \textbf{71.93} \\ \hline
\end{tabular}
\end{center}
\end{table*}

‌\subsection{ITR on two datasets}
Figure.5 illustrates the information transfer rate (ITR) of the different methods for various signal lengths. It is evident that the proposed "incepformer\_ssvep" model performs optimally across all time windows, particularly at shorter signal lengths (0.6s and 0.8s), where the results are particularly notable. Specifically, at a signal length of 0.6s, the model achieves an ITR of 277.33 bits/min, and at 0.8s, the ITR is 265.74 bits/min, which is significantly higher than those of the other methods. This demonstrates the robustness of the proposed approach for short-term signal processing.

\begin{figure*}[ht]
    \centering
    \subfigure[]{
    \begin{minipage}{0.45\linewidth}
        \includegraphics[width=1\textwidth,trim=6cm 5cm 12cm 5cm, clip]{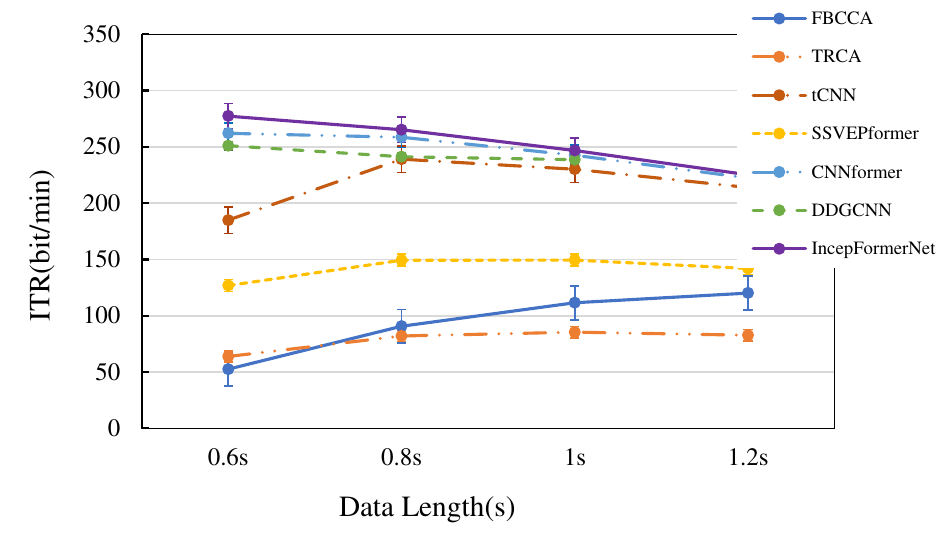}\vspace{10pt}
    \end{minipage}
    }
    \subfigure[]{
    \begin{minipage}{0.45\linewidth}
        \includegraphics[width=1\textwidth,trim=8cm 5cm 10cm 5cm, clip]{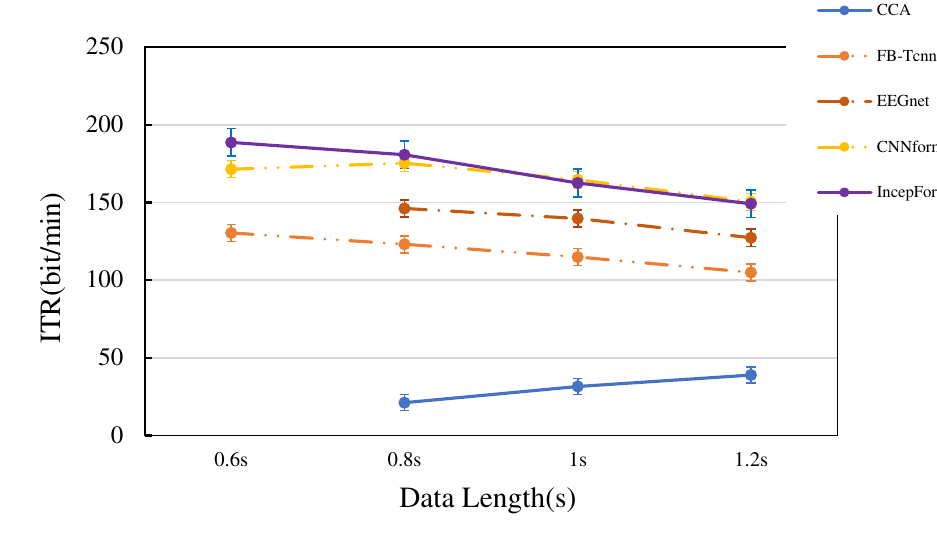}\vspace{10pt}
    \end{minipage}
    }
    \caption{(a)Comparison of average ITR across different methods and time windows on Dataset 1.
    (b)Comparison of average ITR across different methods and time windows on Dataset 2.}
\end{figure*}
As the time window increases, although the ITR slightly declines, the proposed model still maintains a leading position, further validating its stability and effectiveness across different time windows, making it particularly suitable for brain-computer interface systems that require rapid responses. The BETA-ITR table shows the ITR of different models at various time window lengths, and it is clear that the proposed "IncepFormerNet" also performs exceptionally well on the BETA dataset.
Overall, whether in the benchmark or BETA datasets, the proposed model exhibits superior performance across different signal lengths, especially in short-term signal classification tasks, making it well-suited for real-time applications in brain-computer interface systems.

\section{Discussion}
In this study, a new input data representation with rich temporal features are proposed, and a model architecture based on time-domain data is designed, which combines the Inception model's concept with the Transformer model. We then demonstrate the effectiveness of our approach on two public datasets.

‌\subsection{Effectiveness of Inception in Classification Performance}
Inception is the core sub-network structure in the classic GoogLeNet model. It is a sparse matrix that efficiently expresses features\cite{szegedy2015going}. Simply put, Inception assembles multiple convolution or pooling operations into a network module, and when designing neural networks, the entire network structure is built by assembling modules. Its sparse network structure can produce dense data, a feature that improves the network’s performance and ensures the efficient use of computational resources. In the original Inception structure, 1×1 convolutions are used to reduce the dimensionality of the feature maps and the number of model parameters. For SSVEP, because the input data are three-dimensional, the 1×1 convolutions and parallel convolutions with different kernel sizes are replaced with one-dimensional convolutions of various kernel sizes to extract features. This multi-scale convolution operation captures EEG signals at different scales and yields good performance. This approach not only reduces the number of parameters and prevents overfitting but also lowers computational costs, laying the foundation for online SSVEP classification systems.

Figure.6 illustrates the impact of varying the number of blocks on the model's classification accuracy and information transfer rate (ITR). The figure shows the average classification accuracy and standard deviation when using 1 to 6 blocks on the benchmark and BETA datasets. Analysis of the benchmark dataset reveals that as the number of blocks increases, the accuracy rises from 85.17\% to a peak of 87.56\%, followed by a slight decline, reaching 87.01\% with 6 blocks. This indicates that while increasing the number of blocks can enhance classification performance, there is a limit, and too many blocks may lead to a slight decrease in performance.

\begin{figure*}[t]
    \centering
    \subfigure[]{
    \begin{minipage}{0.45\linewidth}
        \includegraphics[width=1\textwidth,trim=5cm 7cm 16cm 3cm,clip]{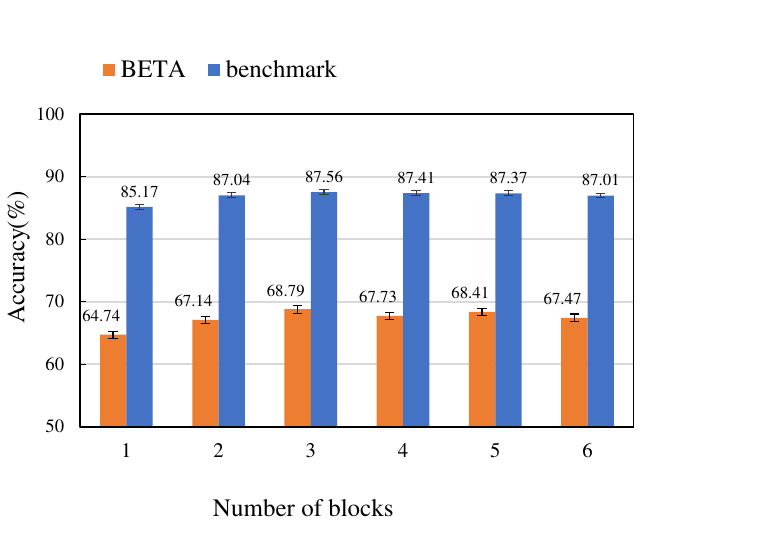}\vspace{10pt}
    \end{minipage}
    }
    \subfigure[]{
    \begin{minipage}{0.45\linewidth}
        \includegraphics[width=1\textwidth,trim=5cm 8cm 16cm 2.7cm,clip]{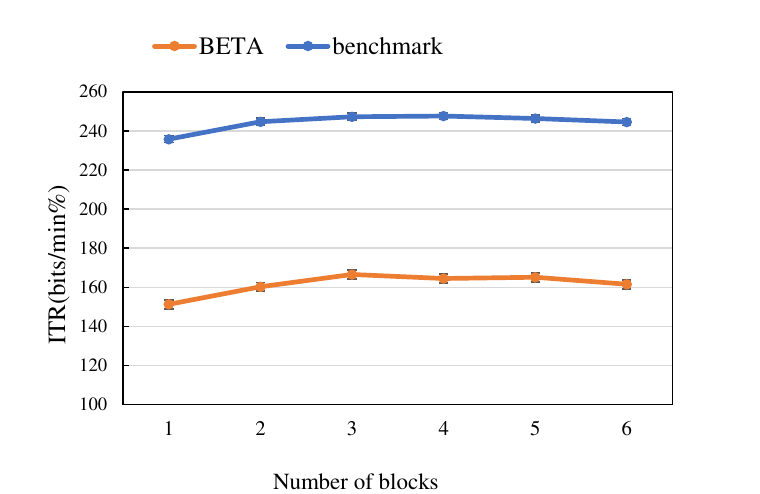}\vspace{10pt}
    \end{minipage}
    }
    \caption{(a)Variations in average accuracy on the Dataset 1 and Dataset 2 using different numbers of blocks.
    (b)Variations in average ITR on the Dataset 1 and Dataset 2 using different numbers of blocks.}
\end{figure*}

The figure displays the trend of ITR as the number of blocks changes, with the ITR reaching its highest point at approximately 248 bits/min when using 3 blocks, followed by a slight decline. Therefore, considering the balance between accuracy and ITR, using 4 blocks as the temporal feature extraction component of the model is the most appropriate choice. When employing 4 blocks, the model demonstrates optimal overall performance. Table.5 shows the selected parameters of the model when using 4 blocks.

\begin{table*}[htbp]
\begin{center}
\caption{Specific parameters of the Inception module used in this study.}
\begin{tabular}{c|c|cc|l}
\hline
module & layer          & \multicolumn{2}{c|}{block}                          & \multicolumn{1}{c}{Explanation} \\ \hline
       &                & \multicolumn{1}{c|}{}              & 1              & kernal\_size=1,padding=‘same’   \\ \cline{4-5} 
       &                & \multicolumn{1}{c|}{}              & 2              & kernal\_size=3,padding=‘same’   \\ \cline{4-5} 
       & dropout=0.5    & \multicolumn{1}{c|}{scale\_blcok}  & 3              & kernal\_size=5,padding=‘same’   \\ \cline{4-5} 
Incep  & Conv1D         & \multicolumn{1}{c|}{}              & 4              & kernal\_size=8,padding=‘same’   \\ \cline{3-5} 
       & BN             & \multicolumn{1}{c|}{}              & max\_pooling×2 & pool\_size=2,padding='same'     \\ \cline{4-5} 
       & Activate='elu' & \multicolumn{1}{c|}{}              & 1              & kernal\_size=32,padding=‘same’  \\ \cline{4-5} 
       &                & \multicolumn{1}{c|}{fusion\_block} & 2              & kernal\_size=16,padding=‘same’  \\ \cline{4-5} 
       &                & \multicolumn{1}{c|}{\textbf{}}     & 3              & kernal\_size=11,padding=‘same’  \\ \hline
\end{tabular}
\end{center}
\end{table*}

The above experiments demonstrated the unique advantages of the Inception module in SSVEP classification. The application of the Inception module in SSVEP classification demonstrates its unique advantages. Through its multi-scale feature extraction capabilities, the Inception module can simultaneously capture local and temporal features in EEG signals, which is critical for identifying different frequencies of SSVEP stimuli. Additionally, its parameter efficiency helps prevent overfitting and makes the model more suitable for real-time applications. In our experiments, the deep network employing the Inception module exhibited high accuracy and low computational cost in the SSVEP classification task and the experimental results further validate the effectiveness of the Inception structure in SSVEP classification tasks. The IncepFormerNet model achieved excellent performance on both the Benchmark and Beta datasets, with accuracy and ITR surpassing those of other models, such as traditional CCA and CNN. This shows that the Inception structure can effectively extract SSVEP features and improve classification performance. Compared with other baselines, this verifies the feasibility of the Inception module in SSVEP classification. Therefore, the Inception module provides an efficient and reliable solution for SSVEP signal classification.

‌\subsection{Attention Mechanism}
The self-attention mechanism allows the model to automatically assign different levels of attention to different parts of the data when processing sequence data\cite{chen2023transformer}, which is useful for capturing key features in EEG signals. In SSVEP classification, EEG signals at different time points may contain information of varying importance. The self-attention mechanism of the Transformer can help the model identify the key signal components related to specific stimulus frequencies, thereby improving classification accuracy. As shown in the ablation experiment table, when only a two-dimensional channel convolution is combined with the multi-head attention mechanism, the accuracy reaches 84\% on Dataset 1 for the classification of 40 different stimuli, which demonstrates the applicability of the Transformer.

The application of the Transformer in our SSVEP classification model, t shows its unique advantages. Through the self-attention mechanism, the Transformer can assign appropriate weights to each time point in the EEG signals, highlighting the key information related to specific stimulus frequencies. Additionally, its ability to model long-range dependencies enables it to better understand the rhythmic characteristics of EEG signals. In our experiments, the model using the Transformer exhibited excellent performance in the SSVEP classification task, validating its feasibility for SSVEP classification. Therefore, the Transformer provides an efficient and powerful tool for classifying of SSVEP signals.

\section{Conclusion}

‌\subsection{Ablation Study}

In ablation experiments, we conducted independent evaluations of the core components of the Incepformer to verify the rationality of its architecture. The experiments were carried out on two datasets, as shown in Figure.7, using a 1-second time window to analyze and compare the performance of the Inception variant and the Transformer structure\cite{ravi2022enhanced}.

\begin{figure}[]
    \centering
    \includegraphics[width=0.5\textwidth,trim=10cm 5cm 8cm 5cm,clip]{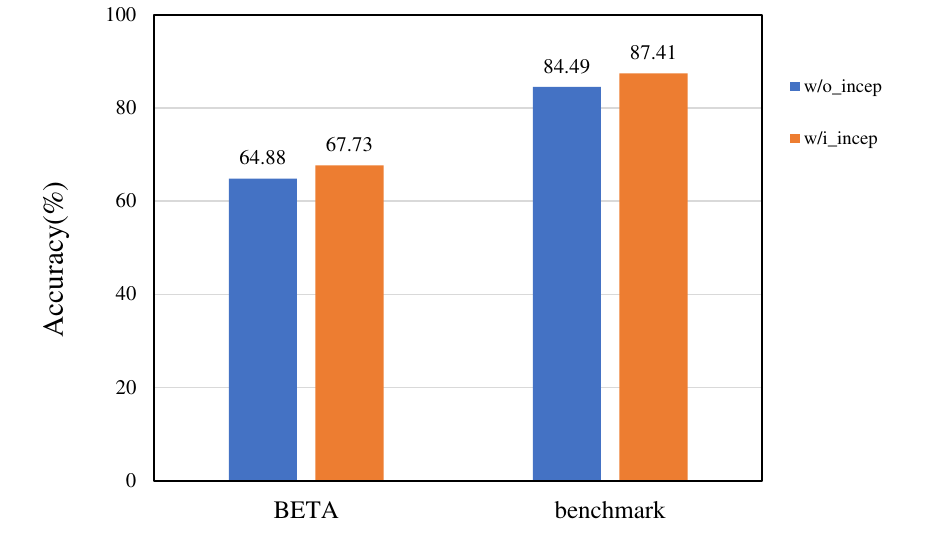}
    \caption{w/o\_incep indicates the absence of the Inception module, whereas w/i\_incep signifies its inclusion.}
\end{figure}

\begin{figure*}[]
    \centering
    \subfigure[]{
    \begin{minipage}{0.3\linewidth}
        \includegraphics[width=1\textwidth]{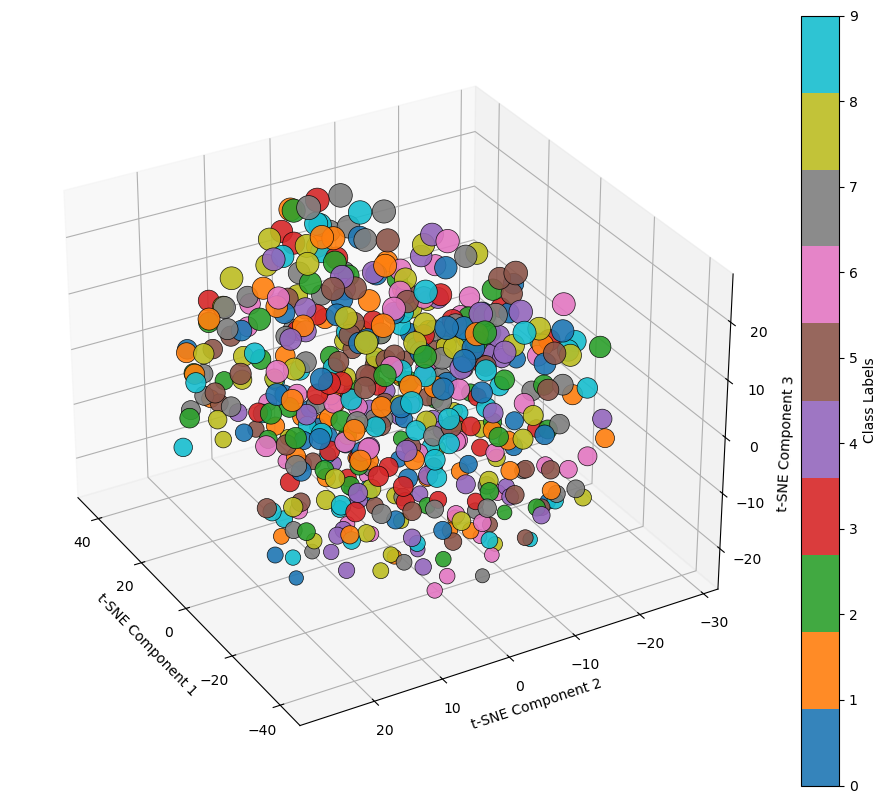}\vspace{10pt}
    \end{minipage}
    }
    \subfigure[]{
    \begin{minipage}{0.3\linewidth}
        \includegraphics[width=1\textwidth]{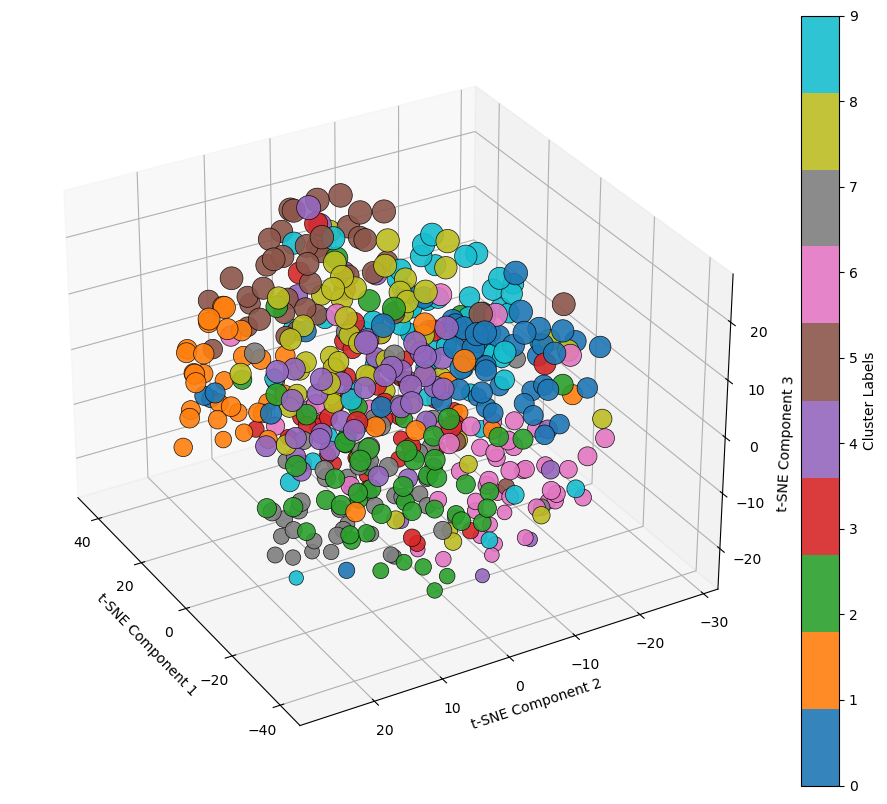}\vspace{10pt}
    \end{minipage}
    }
    \subfigure[]{
    \begin{minipage}{0.3\linewidth}
        \includegraphics[width=1\textwidth]{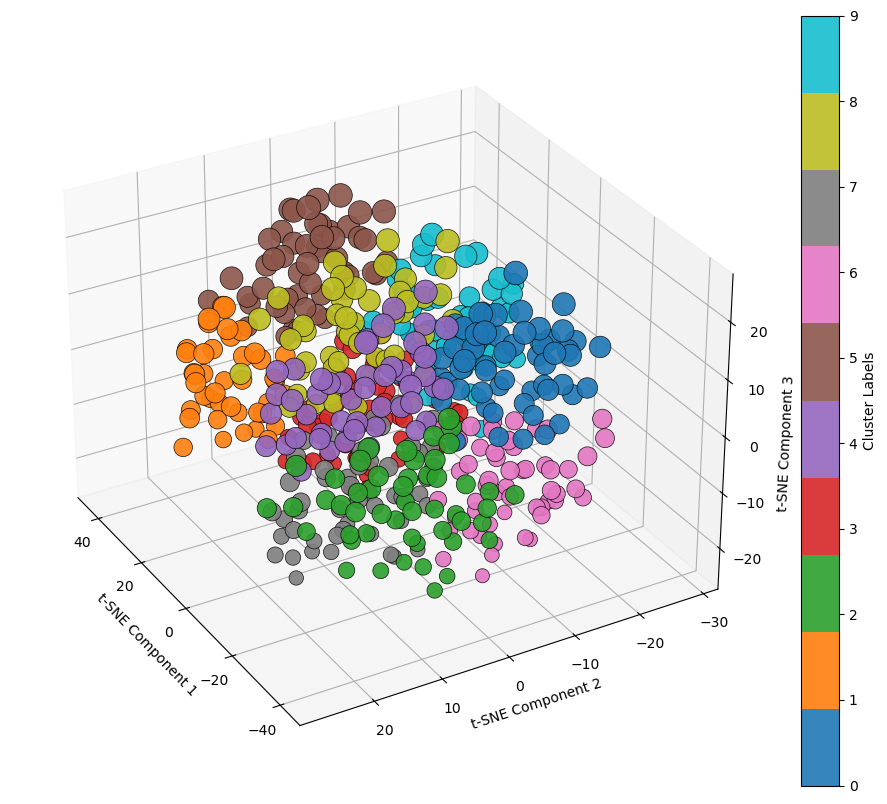}\vspace{10pt}
    \end{minipage}
    }
    \caption{t-SNE visualization results for the first subject on Dataset 1 under different model configurations. (a) Original data (b) With the Inception module (c)Without the Inception module.}
\end{figure*}

First, on Dataset 1, when only using the Transformer structure for learning, the model achieved a classification accuracy of 84.49\% within the 1-second time window. This indicates that even when relying solely on the Transformer’s sequential modeling capability, the model performs well. However, when the Inception variant structure was added, the accuracy increased by nearly 3 percentage points, reaching 87.41\%. This shows that the Inception variant enhances the model's understanding of the 1-second data segments through multi-scale feature extraction, significantly improving classification performance.

On Dataset 2, also using a 1-second time window for analysis, the model achieved an accuracy of 67.73\% when only the Transformer structure was used. After incorporating the Inception variant, the performance also improved by approximately 3 percentage points. This result mirrors the findings on Dataset 1, demonstrating that the Inception module consistently enhances classification accuracy within the 1-second segments across different datasets.

From these ablation experiments, we conclude that the addition of the Inception variant structure significantly strengthens the model’s multi-scale feature extraction capability within 1-second time windows, improving classification performance and achieving near-SOTA results.

\subsection{Feature Visualization}
To further explore why the proposed method achieves better results, we employed t-distributed stochastic neighbor embedding (t-SNE) to visualize three types of features: first, the raw unprocessed data; second, the data features after processing through the model without the Inception structure; and finally, the features extracted through the complete model. We selected data from the first subject in the benchmark dataset and randomly chose ten stimulation frequencies.
As shown in Figure.8 (a) ,the distribution of the raw data is chaotic and lacking any apparent task clustering. In Figure.8 (b), we observe that the data begins to exhibit clustering, although it is still not very pronounced. In Figure.8 (c) the clustering features of the data become significantly clearer, indicating that the features extracted by the proposed model have smaller distances within the same category, thus contributing to improved classification performance.

In this study, we propose a hybrid model based on Inception and Transformer structures, the IncepFormerNet network, which performs well on two public datasets. We validated the accuracy for data lengths of 0.6s, 0.8s, 1.0s, and 1.2s, all of which achieved state-of-the-art (SOTA) performance. This demonstrates that multi-scale extraction of temporal features is both effective and feasible, and also proves the effectiveness of the Transformer-based hybrid model. This model has the potential to promote the practical application of SSVEP-based BCI systems. The IncepFormerNet model ingeniously combines the Inception structure with the Transformer architecture, facilitating multi-scale feature extraction and global information capture for SSVEP signals. This remarkable synergy significantly enhances the decoding accuracy and efficiency of SSVEP-BCI systems. By integrating filter bank techniques, the model effectively leverages the harmonic information inherent in SSVEP signals, further improving classification performance. Its efficient computation makes it particularly well-suited for real-time applications, opening new avenues for research and practical applications in SSVEP-BCI systems. Looking ahead, as advancements in model optimization, multimodal integration, personalized modeling, and online learning continue to evolve, the IncepFormerNet model is poised to have a substantial impact on fields such as healthcare, education, and entertainment, enriching human life with greater convenience and possibilities.
\section*{References}
\bibliographystyle{iopart-num}
\bibliography{main}

\end{document}